\documentclass[11pt]{article}
\usepackage{hyperref}
\usepackage{amssymb}
\usepackage{graphicx}
\usepackage{amsmath}
\usepackage{physics}
\usepackage{authblk}
\usepackage{siunitx}
\usepackage[section]{placeins}
\usepackage{dsfont}
\usepackage[a4paper, total={6.8in, 10in}]{geometry}

\def\be{\begin{equation}}
\def\ee{\end{equation}}
\def\bea{\begin{eqnarray}}
\def\eea{\end{eqnarray}}

\numberwithin{equation}{section}
\title{Quantisation of Klein-Gordon field in $\kappa$ space-time: deformed oscillators and Unruh effect}
\author{E. Harikumar \thanks{harisp@uohyd.ernet.in} and Vishnu Rajagopal \thanks{vishnurajagopal.anayath@gmail.com}}
\affil{School of Physics, University of Hyderabad, \\Central University P.O, Hyderabad-500046, Telangana, India}
\date{}
\begin{document}

\maketitle
\begin{abstract}
In this paper we  study the quantisation of scalar field theory in $\kappa$-deformed space-time.
Using a quantisation scheme that use only field equations, we derive the quantisation rules for deformed 
scalar theory, starting from the $\kappa$-deformed equations of motion. This scheme allows two choices;
(i)a deformed commutation relation between the field and its conjugate which leads to 
usual oscillator algebra, (ii) an undeformed commutation relation between field and its conjugate leading to a
deformed oscillator algebra. This deformed oscillator algebra is used to derive modification to Unruh effect in the 
$\kappa$-space-time.

.
\end{abstract}

\section{Introduction}
Non-commutative space-time was introduced long back as an approach to handle divergences in quantum field 
theory\cite{snyder}. Efforts to construct a renormalisable quantum theory of gravity brought the non-commutative 
space-time into intense scrutiny in last decades\cite{connes}. Appearance of Moyal space-time, a specific type 
of non-commutative space-time, in the low energy limit of a string theory\cite{seiberg-witten} fueled major 
activities in the area of non-commutative theories(for detailed surveys, see the reviews \cite{Nekrasov, Sabo}). 
Non-commutative space-time also emerged naturally in the discussions of different approaches to quantum 
gravity \cite{Glikman, Roberts-Freedenhagen}.

The characteristic features of field theory on non-commutative space-time are their intrinsic non-locality and
non-linearity. They also incorporate a minimal length, in an efficient way into the discussion. This notion of 
minimal length is a common feature of different approaches to microscopic theory of gravity, such as string 
theory\cite{seiberg-witten}, dynamical triangulation\cite{am}, asymptotically safe models, fuzzy physics
\cite{madore, bal} etc. The field theories on non-commutative space-time also show a mixing of UV and IR 
divergences\cite{Minwalla}. The analysis of symmetries of field theories on non-commutative space-time brought 
out the important role played by Hopf algebras\cite{Chaichan, Wess}.

Two widely investigated  non-commutative space-times  are Moyal space-time and $\kappa$-deformed space-time. In 
Moyal space-time\cite{seiberg-witten} , the commutation relation between space-time coordinate is a constant 
tensor. i.e.,
\be
 [\hat{x}_{\mu},\hat{x}_{\nu}]=\Theta_{\mu\nu},
\ee
whereas in $\kappa$ space-time, the space-time coordinates satisfy a Lie algebra type commutation relation
\be
 [\hat{x}_{0},\hat{x}_{i}]=ia\hat{x}_i,~~[\hat{x}_{i},\hat{x}_{j}]=0.\label{kappa-def}
\ee
$\kappa$-deformed space-time appear in the low energy limits of loop gravity models  and it is also the 
space-time associated with Deformed/Doubly Special Relativity(DSR). Since quantum gravity effects are expected to be
important below a fundamental length scale, attempts were made to incorporate such a length scale with principle of relativity, leading to DSR model\cite{Glikman,dsr1,amelio-camelia} and it was shown that the $\kappa$-space-time is the natural arena of DSR\cite{shanmajid}.

Different aspects of field theory models defined on $\kappa$-space-time have been studied in recent times
\cite{lukierski1, lukierski2,  lukierski3, mel1,mel2, mel3,mel4,Paschol,boro,pach}.  In most of these studies one 
start with the deformed dispersion relation, compatible with the $\kappa$-Poincare algebra and arrive at the 
consistent field equations. These field equations  involve higher derivatives terms, signaling the underlying 
non-local nature of these models. Lagrangians are constructed so as to lead to these field equations as 
Euler-Lagrange  equations and hence involve higher derivatives. This makes the canonical quantisation of these 
models a problematic one. Gauge theories in $\kappa$-space-time were constructed and studied in 
\cite{jonke, moller,meyer}.

The canonical quantisation procedure requires the knowledge of the explicit form of the Lagrangian, but there is 
an alternative quantisation method discussed in \cite{ytaka, yt}, which does not require the Lagrangian, rather it uses the equations of motion as the starting point  for quantisation. In \cite{pal} 
the massive spin one gauge field in commutative space-time was quantised using this procedure.  This quantisation method, unlike the 
canonical scheme,  does not use the notion of conjugate momentum. The first step here is to define unequal time 
commutation relation between the field and its adjoint such that the field equations are compatible with 
Heisenberg's equations of motion. This, then leads to commutation relations between the creation and annihilation 
operators that appear in the mode expansion of fields. Apart from the quantisation, this procedure also provides 
an elagant way to construct  conserved currents directly from the equation of motion, without any reference to  
Lagrangian. Using this method one can also obtain the currents corresponding to discrete symmetries \cite{lur}.

The non-commutative field theories are non-local and non-linear, and they also posses higher derivative terms. So 
construction of Lagrangian and quantisation of such non-commutative field theories are not trivial. Hence 
the canonical quantisation for such non-commutative field theories are difficult. In \cite{trg} the quantisation 
of Klein-Gordon field in $\kappa$-deformed space-time was studied and shown that the compatibility between 
quantisation and action of twisted flip operator  \cite{gup} leads to deformed oscillator algebra in 
$\kappa-$Minkowski space-time.

Equations satisfied by different field theory models in $\kappa$-deformed space-time were derived, with out any reference to Lagrangians 
in recent times. In \cite{eh}, Maxwell's equations in $\kappa$-space-time were derived using Feynman's approach
and in \cite{max} the $\kappa$-deformed Maxwell's equations in terms of fields defined in the commuttaive 
space-time were derived by elevating the principle of minimal coupling to non-commutative space-time. 
$\kappa$-deformed geodesic equation was derived by generalising the Feynman's approach to $\kappa$ space-time 
in \cite{geo}. $\kappa$-deformed Dirac equation was analysed in \cite{kDirac}. The approach used in this paper 
can be employed for quantising these models also.

In this paper, we adapt the quantisation scheme of \cite{ytaka,yt, pal} and apply it to $\kappa$-deformed scalar 
field equation, valid up to first order in the deformation parameter. This scalar field equation was derived using 
the quadratic Casimir of the undeformed $\kappa$-Poincare algebra\cite{mel1}. This 
field equation contains infinitely many derivatives and in the commutative limit, reduced to the well known 
Klein-Gordon equation. There is another generalisation of $\kappa$-deformed Klein-Gordon equation, which also 
reduce to the corrrect commutative limit\cite{mel1}. In this paper, we derive the 
unequal time commutation relation between the field and its adjoint, satisfying the modified Klein-Gordon equation. 
Then by appealing to the higher derivative theory methods\cite{Ogs, braga, bol}, we introduce the conjugate momentum 
correspodning to the field and (i) derive the canonical commutation relation between them and show that this leads 
to the usual oscillator algebra between the creation and annihilation operators;(ii) starting with a specific 
deformed algebra satisfied by the creation and annihilation operators, show that the field and conjugate momentum 
satisfy usual commutation relation as in the commutative space-time. We then investigate the modification to 
the Unruh effect\cite{un,lcbc,bir}  due to $\kappa$-deformation using this deformed oscillators. We show that the total 
number of particles seen by an accelerated observer in the Minkowski vacuum is modified due to the 
non-commutativity of the space-time, but the Unruh temperature is unaffected. In \cite{hck,ravi1,ravi2}, Unruh effect in 
$\kappa$-space-time was analysed using different approaches and modifications to Unruh temperature was obtained.

Organisation of this paper is as follows, In Sec.2 we briefly discuss the procedure involved in quantising a 
scalar field using its equation of motion.  We also recall the construction of conserved current  for free fields 
corresponding to the symmetry transformations using this formalism. In Sec.3 we start with $\kappa$-space-time 
whose coordinates satisfy a Lie algebra type commutation relation given in Eq.(\ref{kappa-def}) Using the quadratic 
Casimir of the undeformed $\kappa$-Poincare algebra we set up $\kappa$-deformed Klein-Gordon equation for a particular
choice of realisation, valid upto first order in $a$. Now we follow the quantisation procedure\cite{ytaka,yt} 
outlined in Sec.2 and quantise the Klein-Gordon theory living in $\kappa$-space-time.  Here we find the unequal 
time commutation relation between the field and its adjoint, which is modified due to the space-time 
non-commutativity.  In subsection. 3.1, we derive the modified commutation relation between deformed field and 
its  conjugate momentum and show that the corresponding creation and annihilation operators satisfy the usual 
oscillator algebra.  In subsection 3.2  we take an alternate route. Here we propose a generic, deformed 
oscillator algebra for the deformed creation and annihilate operators and show that this leads to usual 
commutation relation between deformed field and its conjugate momentum as in the commutative space-time. Using 
this deformed oscillator algebra we find that the eigen values of deformed number operator get modified by a 
multiplicative factor. This factor is the same as the one appearing in the modification of the commutation relation 
between deformed creation and deformed annihilate operators. In Sec.4 we derive modifications to  Unruh effect
due to $\kappa$-deformation, by expanding the $\kappa$-deformed Klein-Gordon field in $1 + 1$ dimensions defined in
Minkowski space-time and Rindler space-time, whose frequency modes are connected through Bogoliubov 
transformation.  We show that the Unruh effect gets a modification due to the non-commutativity but the Unruh 
temperature is unaffected. In Sec. 5 we give our concluding remarks. In appendix we show the calculational details
of the construction of deformed conjugate momentum using the formalism of higher derivative theories. Here we work with 
$\eta_{\mu\nu}=$diag$(-1,+1,+1,+1)$.

\section{Quantisation of Klein-Gordon field}
In \cite{yt} Y. Takahashi and H. Umezawa have shown that the quantisation of the fields can be done using their equations of motion. The most interesting part of this procedure is that it does not employ the canonical Lagrange formalism for quantisation. Similarly the conserved currents can also be derived directly from the equations of motion.\cite{ytaka,pal}

According to this scheme, if a free field possess an equation of motion then one can represent it using 
an operator, $\Lambda(\partial)$, which is a polynomial of derivative operators $\partial_{\mu}$ and it is given as
\begin{equation}
 \begin{split}
\Lambda(\partial)&=\sum_{l=0}^{N}\Lambda_{\mu_1\mu_2....\mu_l}\partial^{\mu_1}\partial^{\mu_2}....\partial^{\mu_l}\\
&=\Lambda+\Lambda_{\mu}\partial^{\mu}+\Lambda_{\mu\nu}\partial^{\mu}\partial^{\nu}+\Lambda_{\mu\nu\rho}\partial^{\mu}\partial^{\nu}\partial^{\rho}+.....+\Lambda_{\mu_1\mu_2\mu_3.....\mu_N}\partial^{\mu_1}\partial^{\mu_2}\partial^{\mu_3}.........\partial^{\mu_N}
\end{split}
\end{equation}
This construction is valid when $\Lambda$'s are symmetric in its indices. Using this $\Lambda(\partial)$ operator we express the equations of motion for the field theory under study as
\begin{equation}\label{main}
 \Lambda(\partial)\phi(x)=0.
\end{equation}
Similarly the equation of motion for the adjoint field, $\bar{\phi}(x)$ is given by
\begin{equation}\label{main1}
 \bar{\phi}(x)\Lambda(-\overleftarrow{\partial})=0.
\end{equation} 
In this procedure equation of motion, Eq.(\ref{main}) reduces to Klein-Gordon wave equation
\begin{equation}
\big(\Box-m^2\big)\phi(x)=0,
\end{equation}
by applying an operator called Klein-Gordon divisor, denoted by $d(\partial)$, such that it obeys the relations
\begin{equation}\label{taka}
\Lambda(\partial)d(\partial)=\Box-m^2,  ~~[\Lambda(\partial),d(\partial)]=0
\end{equation}
It is to be emphasised that the operator $d(\partial)$ should have non-vanishing eigen values, i.e, 
det$[d(\partial)]\neq 0$, (see \cite{ytaka} for details). The Eq.(\ref{taka}) is known as 
the $I^{st}$ identity of this quantisation method. For Klein-Gordon field $d(\partial)$ is ${\mathds I}$.

Following \cite{ytaka,yt}, the field operator defined using creation and annihilate operators takes the form
\begin{equation}\label{field}
\phi(x)=\int\frac{d^3p}{\sqrt{(2\pi)^32E_p}}\Big(u_p(x)a(p)+u^*_p(x)a^{\dagger}(p)\Big).
\end{equation}
Here, $u_p(x)$ is a solution of $ \Lambda(\partial)u_p(x)=0$ and $u_p^*(x)$ satisfies
$ u_p^*(x)\Lambda(-\overleftarrow{\partial})=0.$ The creation and annihilate operators appearing in the above mode expansion
 follow canonical commutation relation if the field is bosonic, they follow canonical anti-commutation relation if they are 
fermionic fields. Thus for a Klein-Gordon field the creation and annihilate operators obey the relations
\begin{equation}\label{op}
[a(k),a(k')]=[a^{\dagger}(k),a^{\dagger}(k')]=0,~~ [a(k),a^{\dagger}(k')]=\delta^3(k-k').
\end{equation}
Once the field operator, given in Eq.(\ref{field}) is bosonic (such that their creation and annihilate operators 
obey Eq.(\ref{op})), then they follow an unequal time commutation relation in the light cone, \cite{ytaka,yt}, given by
\begin{equation}\label{co}
[\phi(x),\bar{\phi}(x')]=id(\partial)\Delta(x-x')
\end{equation}
where 
\begin{equation*}
 \Delta(x-x')=\int\frac{d^3p}{(2\pi)^32E_p}\big(e^{-ip(x-x')}-e^{ip(x-x')}\big).
\end{equation*}
Next one constructs the conserved current using the equations of motion \cite{ytaka,yt,lur}. In order to calculate the 
conserved current, one defines an operator $\Gamma_{\mu}(\partial,-\overleftarrow{\partial})$ which is given by
\begin{equation}
\begin{split}
 \Gamma_{\mu}(\partial,-\overleftarrow{\partial})&=\sum_{l=0}^{N-1}\sum_{i=0}^{l}\Lambda_{\mu\mu_1.....\mu_l}\partial_{\mu_1}.....\partial_{\mu_i}(-\overleftarrow{\partial}_{\mu_{i+1}})......(-\overleftarrow{\partial}_{\mu_l})\\
&=\Lambda_{\mu}+\Lambda_{\mu\nu}(\partial^{\nu}-\overleftarrow{\partial}^{\nu})+\Lambda_{\mu\nu\rho}(\partial^{\nu}\partial^{\rho}-\partial^{\nu}\overleftarrow{\partial}^{\rho}+\overleftarrow{\partial}^{\nu}\overleftarrow{\partial}^{\rho})+......
\end{split}
\end{equation}
and it satisfies the identity
\begin{equation}\label{iden}
 (\partial^{\mu}+\overleftarrow{\partial}^{\mu})\Gamma_{\mu}(\partial,-\overleftarrow{\partial})=\Lambda(\partial)-\Lambda(-\overleftarrow{\partial}).
\end{equation}
This identity allows one to construct the conserved currents corresponding to the symmetry of the field equation. 

Now we assume that the field equation and its adjoint are invariant under the transformations
\begin{equation}
 \phi(x)\rightarrow F[x]\textnormal{ and }\bar{\phi}(x)\rightarrow G[x]
\end{equation} 
where $F[x]$ and $G[x]$ are some functionals of the field operator. Now under these transformations Eq.(\ref{main}) and Eq.(\ref{main1}) become
\begin{equation}
 \Lambda(\partial)F[x]=0 \textnormal{ and }G[x]\Lambda(-\overleftarrow{\partial})=0,
\end{equation}
respectively. Now using the identity given in Eq.(\ref{iden}) we get the conserved current as
\begin{equation}\label{conser}
 J_{\mu}(x)=G[x]\Gamma_{\mu}(\partial,-\overleftarrow{\partial})F[x].
\end{equation}
We can easily verify that the covariant derivative of the conserved current, given above, vanishes, i.e, 
$\partial_{\mu}J^{\mu}=0$. 
Thus the Noether current can be derived without the prior knowledge of the exlipict form of the Lagrangian.

For an infinitesimal translation, we have $\delta\phi(x)=-\theta^{\mu}\partial_{\mu}\phi(x)$, and take $F=\delta\phi(x)$ and $G=\bar{\phi}(x)$\cite{lur}. Using the conserved current given in Eq.(\ref{conser}), we construct the conserved 
momentum as
\begin{equation}
 P_{\mu}=\int d^3x \frac{1}{4}\Big(\partial_{\mu}\bar{\phi}(x)\Gamma_0(\partial,-\overleftarrow{\partial})\phi(x)-\bar{\phi}(x)\Gamma_0(\partial,-\overleftarrow{\partial})\partial_{\mu}\phi(x)\Big).
\end{equation}
Only when the field satisfies $i\hbar\partial^\mu\phi(x)=[\phi(x), P^\mu]$ 
were the RHS is evaluated using Eq.(\ref{co}), this quantisation scheme is deemed to be consistent.

\section{$\kappa$-deformed scalar theory and Deformed oscillators} 

In this section we quantise the $\kappa$-deformed Klein-Gordon equation using the formulation discussed in Sec.2 
and obtain a deformed commutation relation between creation and annihilate operators. Using this we evaluate 
the deformed eigen values of creation, annihilate and number operators,  defined in $\kappa$-space-time.
 
The space-time coordinates of $\kappa$-deformed space-time obey Lie-algebra type commutation relations given by
\begin{equation}
[\hat{x}_i,\hat{x}_j]=0,~~[\hat{x}_0,\hat{x}_i]=ia\hat{x}_i
\end{equation}
where $a=\frac{1}{\kappa}$. We choose a specific realisation of $\kappa$-deformed space-time, given by \cite{mel1}
\begin{equation}
 \hat{x}_{\mu}=x_{\alpha}\varphi^{\alpha}_{\mu}.
\end{equation}
Under the undeformed $\kappa$-Poincare algebra, which is the symmetry algebra of the $\kappa$-space-time, the 
derivatives that transform as component of a 4-vector are called Dirac derivatives, and given by
\begin{equation} 
D_{i}=\partial_{i}\frac{e^{-A}}{\varphi},~~
D_{0}=\partial_{0}\frac{\textnormal{sinh}A}{A}-ia\partial_{i} ^2 \frac{e^{-A}}{2\varphi^2}.
\end{equation}
The quadratic Casimir of the undeformed $\kappa$-Poincare algebra is given by
\be
D_{\mu}D^{\mu}= \Box\Big(1+\frac{a^2}{4}\Box\Big),
\ee
where the deformed Laplacian is
\begin{equation}\label{dbox}
\Box=\partial_{i} ^2 \frac{e^{-A}}{\varphi^2}+2\partial_{0} ^2\frac{(1-\textnormal{cosh}A)}{A^2}.
\end{equation}
With a particular choice of realisation, $\varphi=e^{-ap_0}$, we get the $\kappa$-deformed Klein-Gordon equation 
valid upto first non-vanishing term in $a$ as
\begin{equation}
 \Big(\partial_i ^2-\partial_0 ^2-m^2-ia\partial_0\partial_i ^2\Big)\hat{\phi}(x)=0.
\end{equation}
Now we quantise the scalar field obeying this $\kappa$-deformed Klein-Gordon equation, valid upto first order in $a$, using the method summarised in Sec.2. We start by defining
\begin{equation}
 \hat{\Lambda}(\partial)\hat{d}(\partial)=\Big(\Box-m^2-ia\partial_0\partial_i ^2\Big)\hat{d}(\partial)=\Box-m^2-ia\partial_0\partial_i ^2\Big.
\end{equation}
Here we take $\hat{d}(\partial)={\mathds I}$ and we have $\hat{\Lambda}(\partial)=\partial_i ^2-\partial_0 ^2-m^2-ia\partial_0\partial_i ^2$. Thus in this formalism the deformed Klein-Gordon equation is written as 
\begin{equation}
 \hat{\Lambda}(\partial)\hat{\phi}(x)=0
\end{equation}
As in Eq.(\ref{field}) here we decompose the deformed field operator as
\begin{equation}
 \hat{\phi}(x)=\int\frac{d^3p}{\sqrt{(2\pi)^32E_p}}\Big(\hat{u}_p(x)\hat{a}(p)+\hat{u}^*_p(x)\hat{a}^{\dagger}(p)\Big).
\end{equation}
Here $\hat{u}(x)$ appearing in the above deformed wave equation satisfy
\begin{equation}\label{dfield}
 \hat{\Lambda}(\partial)\hat{u}(x)=0.
\end{equation}
Next we find $\hat{u}(x)$ explicitly by solving Eq.(\ref{dfield}), using perturbrative method. Thus we take $\hat{\Lambda}(\partial)$ and $\hat{u}(x)$ to be
\begin{equation}\label{per}
 \hat{\Lambda}(\partial)=\Lambda^{(0)}(\partial)+a\Lambda^{(1)}(\partial),~~\hat{u}(x)=u^{(0)}(x)+a\alpha u^{(1)}(x)
\end{equation}
where $\alpha$ has the dimesion of ${L}^{-1}$. We use Eq.(\ref{per}) in Eq.(\ref{dfield}) and keep the terms valid upto first order in $a$. This gives us two equations corresponding to $a$ independent and $a$ dependent coefficients given by
\begin{equation}
\Lambda^{(0)}(\partial)u^{(0)}(x)=0, ~~
\Lambda^{(1)}(\partial)u^{(0)}(x)+\alpha\Lambda^{(0)}(\partial)u^{(1)}(x)=0
\end{equation}
By solving the first equation we get $u^{(0)}(x)$ as the plane wave solution of the commutative Klein-Gordon 
equation
\begin{equation}
u^{(0)}(x)=e^{-ipx}.
\end{equation}
Substituting this in second equation we get
\begin{equation}\label{green}
\Big(\Box-m^2\Big)u^{(1)}(x)=\frac{1}{\alpha}E_p(\vec{p}_i)^2e^{-ipx},
\end{equation}
where $E_p$, is the commutative energy defined as $E_p=\sqrt{\vec{p}^2+m^2}$. We use Green's function method to solve this inhomogenous differential equation. Using $j(x)=\frac{1}{\alpha}E_p(\vec{p}_i)^2e^{-ipx}$, and the Green's function 
\begin{equation}
\Big(\Box-m^2\Big)G(x-x')=\delta^4(x-x'),
\end{equation}
the solution of Eq.(\ref{green}) is written as 
\begin{equation}
 u^{(1)}(x)=u^{(0)}+\int G(x-x')j(x')d^4x'.
\end{equation}
The Green's function, $G(x-x')$ is given by 
\begin{equation}
\begin{split}
  G(x-x')&=-\int \frac{d^4p}{(2\pi)^4}\frac{1}{p^2+m^2}e^{-ip(x-x')}\\
&=\int \frac{d^4p}{(2\pi)^4}\frac{1}{p_0^2-E^2_p}e^{-ip(x-x')}.
\end{split}
\end{equation}
Here the poles of the integrand are at $p_0=E_p$ and $p_0=-E_p$ respectively. In order to evaluate the integral, we shift the poles by $i\epsilon$, so the above integral becomes
\begin{equation}
\begin{split}
  G(x-x')&=\int \frac{d^4p}{(2\pi)^4}\frac{1}{p_0^2-(E_p-i\epsilon)^2}e^{-ip(x-x')}\\
&=\int\frac{d^3p}{(2\pi)^3}\frac{e^{-i\vec{p}\cdot(\vec{x}-\vec{x}')}}{2E_p}\int \frac{dp_0}{2\pi}e^{ip_0(t-t')}\Big\{\frac{1}{(p_0-E_p)+i\epsilon}-\frac{1}{(p_0+E_p)-i\epsilon}\Big\}\\
&=\int\frac{d^3p}{(2\pi)^3}\frac{e^{-i\vec{p}\cdot(\vec{x}-\vec{x}')}}{2E_p}\Big\{-i\theta(t'-t)e^{iE_p(t-t')}-i\theta(t-t')e^{-iE_p(t-t')}\Big\}\\
&=-\int\frac{d^3p}{(2\pi)^3}\frac{i}{2E_p}\Big\{\theta(t'-t)e^{ip(x-x')}+\theta(t-t')e^{-ip(x-x')}\Big\}
\end{split}
\end{equation}
Using this we calculate $u^{(1)}(x)$ as 
\begin{equation}
\begin{split}
 u^{(1)}(x)&=u^{(0)}(x)+\int d^4x'G(x-x')j(x')\\
&=e^{-ipx}+\int d^4x'G(x-x')\frac{E_p(\vec{p})^2}{\alpha}e^{-ipx'}\\
&=e^{-ipx}-\int d^4x'\int\frac{d^3p'}
{(2\pi)^3}\frac{i}{2E_{p'}}\Big\{\theta(t'-t)e^{ip^\prime(x-x')}+\theta(t-t')e^{-ip^\prime(x-x')}\Big\}\frac{E_p}{\alpha}(\vec{p})^2e^{-ipx'}\\
&=e^{-ipx}-2\pi i\int d^3p\frac{1}{2E_{p'}}\Big\{\theta(t'-t)e^{ip'x}\int\frac{d^4x'}{(2\pi)^4}e^{-i(p'+p)x'}+\theta(t-t')e^{-ip'x}\int\frac{d^4x'}{(2\pi)^4}e^{-i(p-p')x'}\Big\}\frac{E_p}{\alpha}(\vec{p})^2\\
&=\bigg[1-\frac{i\pi}{\alpha}(\vec{p})^2\Big\{-\theta(t'-t)+\theta(t-t')\Big\}\bigg]e^{-ipx}.
\end{split}
\end{equation}
Here the first term on RHS is due to homogenous part and remaining terms come from the inhomogenous part of 
solution. Thus the total solution $\hat{u}(x)$, valid upto first order in $a$, is written as
\begin{equation}\label{u}
 \hat{u}(x)=\bigg[1+a\alpha-ia\pi(\vec{p})^2\Big\{-\theta(t'-t)+\theta(t-t')\Big\}\bigg]e^{-ipx}=\Big(u_p^{(0)}(x)+a\alpha u_p^{(1)}(x)\Big)e^{-ipx}
\end{equation}
Now we denote entire terms in square bracket to be $\widetilde{u}(p)$. It should be noted that at equal times $\widetilde{u}^*(p)=\widetilde{u}(p)$. We see that $\hat{u}(x) \propto u_0(x)$ upto first order in the deformation parameter $a$. Hence the solution to the $\kappa$-deformed 
Klein-Gordon equation valid upto first order in $a$ is 
\begin{equation}\label{wo}
\hat{\phi}(x)=\int\frac{d^3p}{\sqrt{(2\pi)^32E_p}}\Big(\widetilde{u}(p)e^{-ipx}\hat{a}(p)+\widetilde{u}^*(p)e^{ipx}\hat{a}^{\dagger}(p)\Big).
\end{equation}
In the commutative limit, i.e, when $a\rightarrow 0$, $\widetilde{u}(p)$ becomes $1$, we get back the usual 
solution to Klein-Gordon equation. 

Being scalar theory we assume the deformed creation and annihilate operators to have their commutation relations defined in the usual sense as
\begin{equation}\label{crea}
[\hat{a}(k),\hat{a}(k')]=[\hat{a}^{\dagger}(k),\hat{a}^{\dagger}(k')]=0,~~ [\hat{a}(k),\hat{a}^{\dagger}(k')]=\delta^3(k-k').
\end{equation}
Using Eq.(\ref{co}) we consider a $\kappa$-deformed version of this commutation relation as
\begin{equation}\label{comm}
[\hat{\phi}(x),\hat{\bar{\phi}}(x')]=id(\partial)\hat{\Delta}(x-x')=id(\partial)\Big(\Delta(x-x')+af(x-x')\Big)
\end{equation}
where we have assumed the $\hat{\Delta}(x-x')$ to have an $a$ dependent correction term, which is expressed as $\hat{\Delta}(x-x')=\Delta(x-x')+af(x-x')$. 
Hence for a field operator obeying $\kappa$-deformed real Klein-Gordon equation ($\hat{\bar{\phi}}(x')$ becomes $\hat{\phi}(x')$), Eq.(\ref{comm}) becomes
\begin{equation}\label{dcomm}
[\hat{\phi}(x),\hat{\phi}(x')]=i\hat{\Delta}(x-x')=i(\Delta(x-x')+af(x-x'))
\end{equation}
Using the explicit form of the field operator $\hat{\phi}(x)$, i.e, Eq.(\ref{wo}) in Eq.(\ref{dcomm}), we e
valuate $f(x-x')$. We find 
\begin{multline}\label{comm2}
[\hat{\phi}(x),\hat{\phi}(y)]=\int \frac{d^3pd^3p'}{(2\pi)^3\sqrt{2E_p2E_{p'}}}
\Bigg(\Big(u_p^{(0)}(x)u_{p'}^{*(0)}(y)-u_p^{*(0)}(x)u_{p'}^{(0)}(y)\Big)[\hat{a}(p),\hat{a}^{\dagger}(p')]+\\a\alpha\Big(u_p^{(1)}(x)u_{p'}^{*(0)}(y)+u_p^{(0)}(x)u_{p'}^{*(1)}(y)-u_p^{*(0)}(x)u_{p'}^{(1)}(y)-u_p^{*(1)}(x)u_{p'}^{(0)}(y)\Big)[\hat{a}(p),\hat{a}^{\dagger}(p')]\Bigg)
\end{multline}
By using Eq.(\ref{crea}) and using the solution for $u^{(1)}(x)$, this reduces to
\begin{equation}
\begin{split}
 [\hat{\phi}(x),\hat{\phi}(y)]
=i\Delta(x-y)(1+2a\alpha)+ia2\pi\Big(\theta(t'-t)-\theta(t-t')\Big)\int \frac{d^3p}{2E_p(2\pi)^3}\vec{p}^2\Big(e^{-ip(x-y)}-e^{ip(x-y)}\Big)
\end{split}
\end{equation}
Note that last integral vanishes as it is an odd function. We identify the coefficients of the $a$ dependent terms on the RHS as $f(x-y)$, which is giving
\begin{equation}
 f(x-y)=2\alpha\Delta(x-y)
\end{equation}
and Eq.(\ref{dcomm}) becomes
\be
[\hat{\phi}(x),\hat{\phi}(y)]= i(1+2a\alpha)\Delta(x-y)
\ee
\subsection{Case 1}
In this subsection we derive deformed equal time commutation relation between scalar field and its conjugate 
momentum defined in $\kappa$-space-time.

Now we start with the deformed conjugate momentum, defined as $\hat{\pi}(x)=\partial_0{\phi}-ia\partial^2_i\phi$ (see appendix for details), 
which comes from the Ostrogradsky's higher derivative formalism\cite{Ogs,braga,bol}
(note that the approach of \cite{yt} does not use any conjugate momentum definitions.). In the commutative limit, 
i.e., a$\rightarrow 0$, we retrive the commutative conjugate momentum as $\pi(x)=\partial_0\phi(x)$. This 
definition facilitates us to calculate the canonical equal time commutation relation between the field operator 
and its conjugate momentum, valid upto first order in $a$ as
\begin{multline}
[\hat{\phi}(\vec{x},x_0),\hat{\pi}(\vec{y},x_0)]=(\partial_{y_0}-ia\partial_{y_i}^2)\big[\hat{\phi}(x),\hat{\phi}(y)\big]_{x_0=y_0}=i(1+2a\alpha)\frac{\partial}{\partial_{y_0}}\Delta(x-y)\Big|_{x_0=y_0}+a\partial^2_{y_i}\Delta(x-y)\Big|_{x_0=y_0}
\end{multline}
at equal times, i.e, $x_0=y_0$ we have 
\begin{equation}
 \frac{\partial}{\partial_{y_0}}\Delta(x-y)\Big|_{x_0=y_0}=\delta^3(x-y),\textnormal{ and }
 \partial^2_{y_i}\Delta(x-y)\Big|_{x_0=y_0}=0.
\end{equation}
Hence the $\kappa$-deformed canonical equal time commutation relation between field and conjugate momentum takes the form
\begin{equation}
[\hat{\phi}(x,t),\hat{\pi}(y,t)]=i(1+2a\alpha)\delta^3(x-y).
\end{equation}
Here, the creation and annihilation operators satisfy the usual harmonic oscillator algebra(see Eqn.(\ref{crea})).

\subsection{Case 2}
In this subsection we derive an undeformed equal time commutation relation between field and its conjugate 
momentum by proposing a deformed commutator between creation and annihilate operators in $\kappa$-space-time.
 
In the previous subsection, we obtained the deformed commutation relation between field and conjugate momentum by assuming Eq.(\ref{crea}) to be true. Now instead of using Eq.(\ref{crea}), let us assume the creation and annihilate operators to have a $\kappa$-deformation factor in their commutation relation, given as
\begin{equation}\label{acrea}
[\hat{a}(k),\hat{a}(k')]=[\hat{a}^{\dagger}(k),\hat{a}^{\dagger}(k')]=0,~~ [\hat{a}(k),\hat{a}^{\dagger}(k')]=h(a)\delta^3(k-k').
\end{equation}
where $h(a)$ is an arbitrary linear function in $a$, and lim $a\rightarrow 0, h(a)=1$. Now we use this modified commutation relation in Eq.(\ref{comm2}) and follow the above steps to get canonical equal time commutation relation between non-commutative field operator and its conjugate 
momentum as
\begin{equation}\label{xp}
[\hat{\phi}(x,t),\hat{\pi}(y,t)]=i\delta^3(x-y)
\end{equation}
This is obtained for the choice $h(a)=1-2a\alpha$.
 
It is to be noticed that even though the field operator and its conjugate momentum are non-commutative, their commutation relation Eq.(\ref{xp}) does not have any  dependency on non-commutative parameters. 
After substituting $h(a)=1-2a\alpha$ in Eq.(\ref{acrea}) we get the $\kappa$-deformed commutation relation between creation and annihilate operators as
\begin{equation}\label{acrea1}
[\hat{a}(k),\hat{a}(k')]=[\hat{a}^{\dagger}(k),\hat{a}^{\dagger}(k')]=0,~~ [\hat{a}(k),\hat{a}^{\dagger}(k')]=(1-2a\alpha)\delta^3(k-k').
\end{equation}
The conservation law given in Eq.(\ref{conser}) can be used to construct the conserved momentum for the 
field obeying the deformed Klein-Gordon equation. Using this, one can obtain the Hamiltonian 
for the same. Now using the deformed algebra one can verify the compatibility with Heisenberg's equation of motion. 
 
The corresponding deformed vacuum state, $\ket{0}$ is defined as $\hat{a}(p)\ket{0}=0$. Now we define the action of $\hat{a}(p)$ and $\hat{a}^{\dagger}(p)$  on the state $\ket{n}$ as 
\begin{equation}\label{c-}
 \hat{a}(p)\ket{n}=c_-\ket{n-1}
\end{equation}
and
\begin{equation}\label{c+}
 \hat{a}(p)\ket{n}=c_+\ket{n+1}
\end{equation}
respectively. Next we define the $\kappa$-deformed number operator as 
\begin{equation}\label{dn}
 \hat{N}(p)=\hat{a}^{\dagger}(p)\hat{a}(p),
\end{equation}
whose action on state $\ket{n}$ is
\begin{equation}\label{n}
 \hat{N}(p)\ket{n}=g(a)n\ket{n},
\end{equation} 
where $g(a)$ is an arbitrary linear function in $a$, such that in the limit $a\rightarrow 0,g(a)$ becomes $1$. Now we use Eq.(\ref{acrea1}) and Eq.(\ref{n})in Eq.(\ref{c-}) and evaluate $c_-$ to be
\begin{equation}\label{c1}
 c_-=\sqrt{g(a)n}.
\end{equation}
Similarly, using Eq.(\ref{acrea1}) and Eq.(\ref{n}) in Eq.(\ref{c+}), we obtain $c_+$ as
\begin{equation}\label{c2}
 c_+=\sqrt{1+g(a)n-2a\alpha}.
\end{equation}
We write down the expectation value of $\kappa$-deformed number operator using Eq.(\ref{dn}) as
\begin{equation}\label{dn1}
 \bra{n}\hat{N}(p)\ket{n}=\bra{n}\hat{a}^{\dagger}(p)\hat{a}(p)\ket{n}
\end{equation}
Using Eq.(\ref{n}), LHS of above equation is written as
\begin{equation}\label{exp1}
 \bra{n}\hat{N}(p)\ket{n}=g(a)n
\end{equation} 
Similarly using Eq.(\ref{c-}), Eq.(\ref{c+}), Eq.(\ref{c1}) and Eq.(\ref{c2}) in RHS of Eq.(\ref{dn1}), we get
\begin{equation}\label{exp2}
 \bra{n}\hat{a}^{\dagger}(p)\hat{a}(p)\ket{n}=\sqrt{g(a)n}\sqrt{1+(n-1)g(a)-2a\alpha}.
\end{equation}
By comparing Eq.(\ref{exp1}) and Eq.(\ref{exp2}) we get $g(a)=1-2a\alpha$ and thus Eq.(\ref{c-}), Eq.(\ref{c+}) and Eq.(\ref{n}) become
\begin{equation}
 \hat{a}(p)\ket{n}=\sqrt{n(1-2a\alpha)}\ket{n-1}\label{d1}
\end{equation}
\begin{equation}
 \hat{a}^\dagger(p)\ket{n}=\sqrt{(n+1)(1-2a\alpha)}\ket{n+1}\label{d2}
\end{equation}
and
\begin{equation}
 \hat{N}(p)\ket{n}=n(1-2a\alpha)\ket{n}.\label{d3}
\end{equation} 
Here we see that the eigen values of all the operators $\hat{a}(p),\hat{a}^{\dagger}(p)$ and $\hat{N}(p)$ get 
modified by a multiplicative factor $(1-2a\alpha)$ under $\kappa$-deformation. This modification is due to the 
deformed oscillator algebra. We recover the commutative eigen values of the above mentioned operators in the 
limit $a\rightarrow 0$.

\section{Deformed Unruh effect}
In this section we use the deformed oscillator algebra given in Eq.(\ref{acrea1}) and study the modifications 
in Unruh effect due to $\kappa$-deformation.

A uniformly accelerating (with constant proper acceleration $A$) observer in Minkowski space-time, observes the 
particles in a thermal bath with temperature, $T=\frac{A\hbar}{2\pi k}$ in the Minkowski vacuum 
and this is known as Unruh effect \cite{un,lcbc,bir}

In \cite{hck} $a$ dependent correction term to the Unruh effect was obtained by considering the interaction
between 
detector and scalar field defined in $\kappa$-space-time. In \cite{ravi1} correction to Unruh effect, 
valid up to order $a^2$  that comes from the response function of a uniformly accelerating detector 
coupled to $\kappa$-deformed Klein-Gordon field written in commutative space-time was discussed. 
Similarly in \cite{ravi2} $a$ dependent correction to the Unruh effect was calculated by analysing the 
response function of a uniformly accelerating detector coupled to massless $\kappa$-deformed Dirac field.

Here we use an alternative approach used in the commutative space-time in calculating the Unruh efect. This method 
uses the Bogoliubov transformations, relating the creation and annihilation operators appearing in the mode 
expansion of fields in two different basis. Using this relation, one calculate the vacuum expectation value of 
the number operator defined in one basis(by accelerating observer), over the vacuum defined in the second basis
(by a stationary observer). This expectation value is shown to be non-zero and in the commutative space-time, the 
value obtained shows the existence of particle in thermal bath with a temperature related to the constant 
acceleration of the observer, in the vacuuum defined by the stationary observer in the Minkowski space-time. 
We use this approach to calculate the modification to Unruh effect in the $\kappa$-deformed space-time.

 We expand the $\kappa$-deformed scalar field in $1+1$ dimensions defined in two different basis, namely Minkowski 
 space-time and Rindler space-time. Then we use Bogoliubov transformation to connect the frequency modes of the 
 left moving sectors of Minkowski space-time with that of the Rindler space-time. Now use this Bogoliubov 
 coefficients and the deformed oscillator algebra to derive the $\kappa$-deformed modications to the Unruh 
 effect.

In $1+1$ dimensions the Minkowski metric takes the form
\begin{equation}\label{met1}
 ds^2=-dt^2+dz^2
\end{equation}
By inspecting the line element itself we notice that the metric is static. The metric has both the Right Rindler Wedge (RRW) and Left Rindler Wedge (LRW) defined by the regions $|t|<z$ and $|t|<-z$ respectively. Now in RRW, we do a coordinate transformation defined by \cite{lcbc,bir}
\begin{equation}\label{tz1}
 t=\frac{e^{A\zeta}}{A}\textnormal{sinh}A\tau,~~z=\frac{e^{A\zeta}}{A}\textnormal{cosh}A\tau
\end{equation}
where $A$ is a positive constant. Thus the Minkowski line element defined in Eq.(\ref{met1}) takes the form
\begin{equation}\label{met2}
 ds^2=-e^{2A\zeta}(dt^2-dz^2).
\end{equation}
Next we consider the massless deformed scalar field satisfying the deformed Klein-Gordon equation in $1+1$ dimension given by
\begin{equation}
 (\partial^2_{z}-\partial^2_t-ia\partial_t\partial^2_z)\hat{\phi}(x)=0
\end{equation}
The deformed field operator is decomposed in $\kappa$-Minkowski space-time as 
\begin{equation}\label{dmin}
 \hat{\phi}(z,t)=\int\frac{dk}{\sqrt{4\pi k}}\Big(\hat{b}_{-k}e^{-ik(t-z)}\widetilde{u}(k)+\hat{b}_{+k}e^{-ik(t+z)}\widetilde{u}(k)+\hat{b}^{\dagger}_{-k}e^{ik(t-z)}\widetilde{u}^*(k)+\hat{b}^{\dagger}_{+k}e^{ik(t+z)}\widetilde{u}^*(k)\Big)
\end{equation}
where the Minkowski vacuum state $\ket{0}_M$ is defined as $\hat{b}_{+k}\ket{0}_M=0$ and $\hat{b}_{-k}\ket{0}_M=0$ and the deformed commutation relation is given by 
\begin{equation}
 [\hat{b}_{\pm k},\hat{b}_{\pm k'}^{\dagger}]=(1-2a\alpha)\delta(k-k')
\end{equation}
and all other commutators vanish. Now we define the coordinates in terms of the light cone variables
\begin{equation}\label{uv}
 U=t-z,~~V=t+z
\end{equation}
and we write the deformed field operator as a sum of left and right moving parts as
\begin{equation}
 \hat{\phi}(t,z)=\hat{\phi}_+(V)+\hat{\phi}_-(U).
\end{equation}
Here $\hat{\phi}_+(V)$ and $\hat{\phi}_-(U)$ represents the left and right moving sectors of the deformed field in $\kappa$-Minkowski space-time, respectively. Since $\hat{\phi}_-(U)$ and $\hat{\phi}_+(V)$ do not interact each other, we consider only the left moving sector, $\hat{\phi}_+(V)$, i.e,
\begin{equation}
 \hat{\phi}_+(V)=\int\frac{dk}{\sqrt{4\pi k}}\Big(\hat{b}_{+k}e^{-ikV}\widetilde{u}(k)+\hat{b}^{\dagger}_{+k}e^{ikV}\widetilde{u}^*(k)\Big).
\end{equation}
From Eq.(\ref{met1}) and Eq.(\ref{met2}), we see that theses metrics are conformally invariant in $1+1$ dimensions. Therefore the solution corresponding to the deformed 
Klein-Gordon equation in RRW has the form given by
\begin{equation}\label{left}
 \hat{\phi}_+(V)=\int\frac{dw}{\sqrt{4\pi w}}\Big(\hat{a}^R_{+w}e^{-iwv}\widetilde{u}(w)+\hat{a}^{\dagger R}_{+w}e^{iwv}\widetilde{u}^*(w)\Big)=\int dw\Big(\hat{a}^R_{+w}g_w(v)\widetilde{u}(w)+\hat{a}^{\dagger R}_{+w}g^*_w(v)\widetilde{u}^*(w)\Big)
\end{equation}
where $g_w(v)=\frac{e^{-iwv}}{\sqrt{4\pi w}}, v=\tau+\zeta$ and $u=\tau-\zeta$. Thus using Eq.(\ref{tz1}) and Eq.(\ref{uv}), $U$ and $V$ becomes 
\begin{equation}\label{uv1}
 U=\frac{-e^{-Au}}{A},~~V=\frac{e^{Av}}{A}
\end{equation}
Now we define $\widetilde{g}_w(v)=g_w(v)\widetilde{u}(w)$ and reexpress Eq.(\ref{left}) as
\begin{equation}
 \hat{\phi}_+(V)=\int dw\Big(\hat{a}^R_{+w}\widetilde{g}_w(v)+\hat{a}^{\dagger R}_{+w}\widetilde{g}_w^*(v)\Big).
\end{equation}
Here $\hat{a}^R_{+w}(k)$, $\hat{a}^{\dagger R}_{+w}(k)$ satisfy deformed algebra (given in Eq.(\ref{acrea1})),
\begin{equation}\label{ar}
 [\hat{a}^R_{+w}(k),\hat{a}^{\dagger R}_{+w}(k')]=(1-2a\alpha)\delta(k-k')
\end{equation}
and all other commutators vanish. Similarly one can express the left moving sector of deformed field operator 
$\hat{\phi}_+(V)$ in LRW with the condition $V<0<U$. Using LRW coordinates $(\bar{\tau},\bar{\zeta})$ defined by 
\begin{equation}\label{tz2}
 t=\frac{e^{A\bar{\zeta}}}{A}\textnormal{sinh}A\bar{\tau},~~
 z=\frac{e^{A\bar{\zeta}}}{A}\textnormal{cosh}A\bar{\tau},
\end{equation}
and with the definitions $\bar{v}=\bar{\tau}-\bar{\zeta}$ and $\bar{u}=\bar{\tau}+\bar{\zeta}$, 
$U$ and $V$ become
\begin{equation}
 U=\frac{e^{A\bar{u}}}{A},~~V=-\frac{e^{-A\bar{v}}}{A}
\end{equation}
Hence in LRW, the deformed field operator takes the form
\begin{equation}
 \hat{\phi}_+(V)=\int dw\Big(\hat{a}^L_{+w}\widetilde{g}_w(\bar{v})+\hat{a}^{\dagger L}_{+w}\widetilde{g}_w^*(\bar{v})\Big)
\end{equation}
where the creation and annihilate operators in LRW satisfy the deformed algebra (given in Eq.(\ref{acrea1}))
\begin{equation}\label{al}
 [\hat{a}^L_{+w}(k),\hat{a}^{\dagger L}_{+w}(k')]=(1-2a\alpha)\delta(k-k')
\end{equation}
and all other commutation relations vanish. The static vacuum state in RRW and LRW, i.e., the Rindler vacuum, $\ket{0}_R$, is defined by $\hat{a}^R_{+w}\ket{0}_R=\hat{a}^L_{+w}\ket{0}_R=0$.

We use Bogoliubov coefficients $\alpha^R_{wk},\beta^R_{wk},\alpha^L_{wk}$ and $\beta^L_{wk}$, to connect the coefficients of the creation and annihilate operators in Minkowski space-time with the Right and Left Rindler Wedges given by \cite{lcbc, bir}
\begin{equation}\label{bg1}
 \Theta(V)\widetilde{g}_w(v)=\int\frac{dk}{\sqrt{4\pi k}}(\alpha^R_{wk}e^{-ikV}+\beta^R_{wk}e^{ikV})
\end{equation}
\begin{equation}\label{bg2}
 \Theta(-V)\widetilde{g}_w(\bar{v})=\int\frac{dk}{\sqrt{4\pi k}}(\alpha^L_{wk}e^{-ikV}+\beta^L_{wk}e^{ikV})
\end{equation}
where $\Theta(V)$ is the usual step function. We first evaluate $\alpha^R_{wk}$. We multiply Eq.(\ref{bg1}) by $\frac{e^{ikV}}{2\pi}$ for $k>0$ and integrating over $V$ and we obtain,
\begin{equation}
\begin{split}
 \alpha^R_{wk}=\sqrt{4\pi k}\int \frac{dV}{2\pi}e^{ikV}\widetilde{g}_w(v)
\end{split}
\end{equation}
Now using Eq.(\ref{uv1}) this is rewritten as
\begin{equation}
 \alpha^R_{wk}=\frac{1}{2\pi}\sqrt{\frac{k}{w}}\widetilde{u}(w)\int dV (AV)^{-\frac{iw}{A}}e^{ikV}
\end{equation}
Next we do a change of variable, $V=\frac{ix}{k}$ and using the Gamma function we get 
\begin{equation}
\begin{split}
 \alpha^R_{wk}=\frac{1}{2\pi}\sqrt{\frac{k}{w}}\widetilde{u}(w)\int \frac{idx}{k}\Big(A\frac{ix}{k}\Big)^{-\frac{iw}{A}}e^{-x}\\
=\frac{ie^{\frac{\pi w}{2A}}}{2\pi\sqrt{wk}}\widetilde{u}(w)\Big(\frac{A}{k}\Big)^{-\frac{iw}{A}}\Gamma\Big(1-\frac{iw}{A}\Big)
\end{split}
\end{equation}
We find $\beta^R_{wk}$  by multiplying Eq.(\ref{bg1}) by $e^{-ikV}$ and following the above procedure and by changing the variable to $V=-\frac{ix}{k}$, $\beta^R_{wk}$ as
\begin{equation}
 \beta^R_{wk}=-\frac{ie^{-\frac{\pi w}{2A}}}{2\pi\sqrt{wk}}\widetilde{u}(w)\Big(\frac{A}{k}\Big)^{-\frac{iw}{A}}\Gamma\Big(1-\frac{iw}{A}\Big)
\end{equation}
Similarly, repeating the above procedures using Eq.(\ref{bg2}), we get $\alpha^L_{wk}$ and $\beta^L_{wk}$ as
\begin{equation}
 \alpha^L_{wk}=-\frac{ie^{\frac{\pi w}{2A}}}{2\pi\sqrt{wk}}\widetilde{u}(w)\Big(\frac{A}{k}\Big)^{\frac{iw}{A}}\Gamma\Big(1+\frac{iw}{A}\Big),~~\beta^L_{wk}=\frac{ie^{-\frac{\pi w}{2A}}}{2\pi\sqrt{wk}}\widetilde{u}(w)\Big(\frac{A}{k}\Big)^{\frac{iw}{A}}\Gamma\Big(1+\frac{iw}{A}\Big).
\end{equation}
On inspecting the above relations we find the Bogoliubov coefficients satisfy the relations given by
\begin{equation}
 \beta^L_{wk}=-e^{-\frac{\pi w}{A}}\alpha^{R*}_{wk},~~\beta^R_{wk}=-e^{-\frac{\pi w}{A}}\alpha^{L*}_{wk}
\end{equation}
By substituting the above relations in Eq.(\ref{bg1}) and Eq.(\ref{bg2}) and after grouping them as positive and 
negative frequency modes, we observe that the functions defined by 
 $G_w(V)$ and $\bar{G}_w(V)$ are the coefficients of the positive frequency modes, $e^{-ikV}$ in Minkowski space-time given by Eq.(\ref{dmin}). In this way we find
\begin{equation}
 G_w(V)=\Theta(V)\widetilde{g}_w(v)+\Theta(-V)\widetilde{g}^*_w(\bar{v})e^{-\frac{\pi w}{A}},
\end{equation}
\begin{equation}
 \bar{G}_w(V)=\Theta(V)\widetilde{g}_w(\bar{v})+\Theta(-V)\widetilde{g}^*_w(v)e^{-\frac{\pi w}{A}},
\end{equation}
and thus we get
\begin{equation}
 \Theta(V) \widetilde{g}_w(v) \propto G_w(V)-\bar{G}^*_w(V)e^{-\frac{\pi w}{A}}
\end{equation}
\begin{equation}
 \Theta(-V) \widetilde{g}_w(\bar{v}) \propto \bar{G}_w(V)-G^*_w(V)e^{-\frac{\pi w}{A}}
\end{equation} 
After substituting these equations into the mode decomposition of the field in the Rindler space-time, i.e.,
\begin{equation}
 \hat{\phi}_+(V)=\int dw\Big(\Theta(V)\big(\hat{a}^R_{+w}\widetilde{g}_w(v)+\hat{a}^{\dagger R}_{+w}\widetilde{g}^*_w(v)\big)+\Theta(-V)\big(\hat{a}^L_{+w}\widetilde{g}_w(\bar{v})+\hat{a}^{\dagger L}_{+w}\widetilde{g}^*_w(\bar{v})\big)\Big),
\end{equation}
we find that the integrand is proportional to
\begin{equation}
 G_w(V)\Big(\hat{a}^R_{+w}-e^{-\frac{\pi w}{A}}\hat{a}^{\dagger L}_{+w}\Big)+\bar{G}_w(V)\Big(\hat{a}^L_{+w}-e^{-\frac{\pi w}{A}}\hat{a}^{\dagger R}_{+w}\Big)+G^*_w(V)\Big(\hat{a}^{\dagger R}_{+w}-e^{-\frac{\pi w}{A}}\hat{a}^{L}_{+w}\Big)+\bar{G}^*_w(V)\Big(\hat{a}^{\dagger L}_{+w}-e^{-\frac{\pi w}{A}}\hat{a}^{R}_{+w}\Big)
\end{equation}
Here $G_w(V)$ and $\bar{G}_w(V)$ are the co-efficients of the positive frequency modes in Minkowski space-time, so the operators 
$\hat{a}^R_{+w}-e^{-\frac{\pi w}{A}}\hat{a}^{\dagger L}_{+w}$ and $\hat{a}^L_{+w}-e^{-\frac{\pi w}{A}}\hat{a}^{\dagger R}_{+w}$ annihilate the Minkowski vacuum, i.e., 
\begin{equation}\label{min1}
 \Big(\hat{a}^L_{+w}-e^{-\frac{\pi w}{A}}\hat{a}^{\dagger R}_{+w}\Big)\ket{0}_M=0
\end{equation}
\begin{equation}\label{min2}
 \Big(\hat{a}^R_{+w}-e^{-\frac{\pi w}{A}}\hat{a}^{\dagger L}_{+w}\Big)\ket{0}_M=0
\end{equation}
where creation and annihilate operators in the RRW and LRW obey the commutation relations given by Eq.(\ref{ar}) and Eq.(\ref{al}), respectively. By using Eq.(\ref{ar}) and Eq.(\ref{al}) in Eq.(\ref{min1}) and Eq.(\ref{min2}) we find
\begin{equation}
 {}_M\bra{0}\hat{a}^{\dagger R}_{+w}\hat{a}^R_{+w}\ket{0}_M=e^{-\frac{2\pi w}{A}}(1-2a\alpha)+e^{-\frac{2\pi w}{A}}{}_M\bra{0}\hat{a}^{\dagger L}_{+w}\hat{a}^L_{+w}\ket{0}_M
\end{equation}
\begin{equation}
 {}_M\bra{0}\hat{a}^{\dagger L}_{+w}\hat{a}^L_{+w}\ket{0}_M=e^{-\frac{2\pi w}{A}}(1-2a\alpha)+e^{-\frac{2\pi w}{A}}{}_M\bra{0}\hat{a}^{\dagger R}_{+w}\hat{a}^R_{+w}\ket{0}_M
\end{equation}
We solve this equations simultaneously and we obtain
\begin{equation}\label{dunruh}
 {}_{M}\bra{0}\hat{a}^{\dagger L}_{+w}\hat{a}^L_{+w}\ket{0}_M={}_{M}\bra{0}\hat{a}^{\dagger R}_{+w}\hat{a}^R_{+w}\ket{0}_M=\frac{1-2a\alpha}{e^{\frac{2\pi w}{A}}-1}.
\end{equation}
Here we notice that the vacuum expection value of the deformed number operator for the Rindler particles get
modified by a $(1-2a\alpha)$ factor. The deformation in the vacuum expectation value of the Rindler particle 
number operator is due to the deformed oscillator algebra derived in Eq.(\ref{acrea1}). Note that in the 
commutative limit the $a$ dependent correction in Eq.(\ref{dunruh}) vanishes, reducing it to the commutative 
value. In $\cite{hck,ravi1,ravi2}$ also, corrections to the Unruh effect were obtained. But in 
these cases the $\kappa$-deformation leads to correction in Unruh temperature unlike in the present case. Since 
$\hat{u}(x)\propto u_0(x)$ (see Eq.(\ref{u})) upto first order in $a$, we note that the frequency, $w$ is not 
modified. This may not be true if corrections to all orders in $a$ are included in the calculation. Since f
requency is not modified upto first order in $a$, from Eq.(\ref{dunruh}) we see that Unruh temperature is 
unaffected. 

\section{Conclusions}

In this paper we have quantised the $\kappa$-deformed scalar field, which staisfies deformed Klein-Gordon equation in the 
$\kappa$-space-time,  for a particular choice of 
realisation. This equation is construced from the quadratic Casimir of the corresponding symmetry algebra, viz;
undeformed $\kappa$-Poincare algebra.  

In the $\kappa$-space-time, different possible generalisations of Klein-Gordon field equations, all of which 
reduced to correct commutative limit have been studied. Construction of field equation is guided by the 
conditions (i) that they all, in the limit of vanishing non-commutative parameter, give well established 
equations of commutative space-time,(ii) they all should give same deformed energy-momentum relations valid 
for the non-commutative space-time. In the case of $\kappa$-deformed space-time, Maxwell's equations\cite{eh, max}, geodesic 
equation\cite{geo} and Dirac equation \cite{kDirac} were constructed without any reference to Lagrangian.

Starting from the deformed equation of motion for the scalar field theory, after constructing the operators
$\Lambda(\partial)$, Klein-Gordon divisor and $\Gamma_\mu$ operator, we have derived, deformed, unequal 
time commutation relation between field and its adjoint, valid upto first order in the deformation parameter 
$a$. We then adopted the definition of (deformed) conjugate momentum given by the formalism of higher
derivative theories\cite{Ogs,braga,bol} and derived equal time commutation relations between the field and its
conjugate momentum. In subsection. 3.1 we have derived a deformed 
commutation relation between scalar field and its conjugate momentum in $\kappa$-space-time, valid upto first order in
$a$. This leads to usual harmonic oscillator algebra between the deformed creation and annihilate 
operators. In subsection. 3.2 we showed that it is possible to obtain deformed commutation relations 
between creation and annihilation operators. Using this set of deformed operators, we have constructed number 
basis and showed that under $\kappa$-deformation, eigen values of the creation, annihilate and number operators 
pick up a $(1-2a\alpha)$ factor. This factor is the same as the $\kappa$-deformed modification present in the 
deformed oscillator algebra. Similar modification in the eigen values were seen in the context of $q$-deformed 
oscillators also \cite{qdef}.

In \cite{trg} quantisation of the scalar field satisfying $\kappa$-deformed Klein-Gordon equation given by $(\Box-m^2)\hat{\phi}(x)=0$, was studied. The compatibility between the flip operator and the modified product rule resulted in the modification to the commutation relation between scalar fields defined at two different points. These modified commutation relation was used to derive the deformed oscillator algebra. In the present case product rule is not modified under $\kappa$-deformation, but the unequal time commutation relation between field and its adjoint is modified. This resulted in the deformed oscillator algebra (which is different from one obtained in \cite{trg}).

We used the deformed oscillator algebra derived to study the modification in the Unruh effect, due to 
$\kappa$-deformation of the space-time.  We employed the approach of Bogoliubov transformations in deriving 
the modification to Unruh effect. We showed that the vacuum expectation value of the deformed number operator defined by the Rindler observer
gets modified by a factor which is same as the one that appeared in the deformed oscillator algebra. 
Interestingly, this modification does not alter the Unruh temperature  and it remains unchanged under 
$\kappa$-deformation.

\subsection*{Appendix}
In this appendix, we summarise the definition of conjugate momenta for Lagrangian having higher derivatives. Then 
using this, we derive the conjugate momenta valid upto first order in the deformation parameter for the 
$\kappa$-deformed scalar theory.

We consider the Lagrangian of form $\mathcal{L}(\phi,\partial_{\mu}\phi,\partial_{\mu}\partial_{\nu}\phi)$ and vary the action as 
\begin{equation*}
 \delta S=\delta \int_{t_1}^{t_2} dt\int d^3x \mathcal{L}(\phi,\partial_{\mu}\phi,\partial_{\mu}\partial_{\nu}\phi)=0
\end{equation*}
such that $\delta\phi(t_1)=\delta\phi(t_2)=0,~~\delta\dot{\phi(t_1)}=\delta\dot{\phi(t_2)}=0$ \cite{braga}. 
\begin{multline*}
 \implies\delta S=\delta \int_{t_1}^{t_2} dt\int d^3x\bigg\{\bigg(\frac{\partial\mathcal{L}}{\partial\phi}-\partial_{\mu}\Big(\frac{\partial\mathcal{L}}{\partial(\partial_{\mu}\phi)}\Big)+\partial_{\mu}\partial_{\nu}\Big(\frac{\partial\mathcal{L}}{\partial(\partial_{\mu}\partial_{\nu}\phi)}\Big)\bigg)\delta\phi+\\\partial_{\mu}\bigg(\frac{\partial\mathcal{L}}{\partial(\partial_{\mu}\phi)}\delta\phi-\partial_{\nu}\Big(\frac{\partial\mathcal{L}}{\partial(\partial_{\mu}\partial_{\nu}\phi)}\Big)\delta\phi+\frac{\partial\mathcal{L}}{\partial(\partial_{\mu}\partial_{\nu}\phi)}\delta(\partial_{\nu}\phi)\bigg)\bigg\}
\end{multline*}
where, the Euler-Lagrange equation of motion is \cite{bol}
\begin{equation*}
 \frac{\partial\mathcal{L}}{\partial\phi}-\partial_{\mu}\bigg(\frac{\partial\mathcal{L}}{\partial(\partial_{\mu}\phi)}\bigg)+\partial_{\mu}\partial_{\nu}\bigg(\frac{\partial\mathcal{L}}{\partial(\partial_{\mu}\partial_{\nu}\phi)}\bigg)=0.
\end{equation*}
Thus on simplyfing we get
\begin{equation*}
 \delta S= \int d^3x\bigg\{\bigg(\frac{\partial\mathcal{L}}{\partial(\partial_0\phi)}+\partial_0\Big(\frac{\partial\mathcal{L}}{\partial(\partial_0\partial_0\phi)}\Big)-2\partial_i\Big(\frac{\partial\mathcal{L}}{\partial(\partial_i\partial_0\phi)}\Big)\bigg)\delta\phi-\frac{\partial\mathcal{L}}{\partial(\partial_0\partial_0\phi)}\delta(\partial_0\phi)\bigg\}.
\end{equation*}
We define conjugate momenta as
\begin{equation*}
 \pi^{(1)}=-\frac{\partial\mathcal{L}}{\partial(\partial_0\phi)}-\partial_0\Big(\frac{\partial\mathcal{L}}{\partial(\partial_0\partial_0\phi)}\Big)+2\partial_i\Big(\frac{\partial\mathcal{L}}{\partial(\partial_i\partial_0\phi)}\Big)
\end{equation*}
and
\begin{equation*}
 \pi^{(2)}=\frac{\partial\mathcal{L}}{\partial(\partial_0\partial_0\phi)}
\end{equation*}
respectively. Now the Lagrangian of $\kappa$-deformed Klein-Gordon field is written using Dirac derivatives as
\begin{equation*}
 \mathcal{L}=\frac{1}{2}D^{\mu}\phi D_{\mu}\phi+\frac{1}{2}m^2\phi^2.
\end{equation*}
Using the components of Dirac derivative, above Lagrangian, valid upto first order in $a$ becomes
\begin{equation*}
 \mathcal{L}=\frac{1}{2}\partial_{\mu}\phi\partial^{\mu}\phi+\frac{1}{2}m^2\phi^2-ia\partial_{\mu}\phi\partial_{\mu}\partial_{\lambda}\phi\delta^{\mu 0}\delta^{\nu i}\delta^{\lambda i}.
\end{equation*}
Using the above definition we find conjugate momenta  as
\begin{equation}
 \pi^{(1)}=\partial_0\phi-ia\partial^2_i\phi,~~\pi^{(2)}=0.
\end{equation}

\subsection*{\bf Acknowledgments}
EH thanks Prof. V. Srinivasan for introducing to the reference \cite{ytaka}.
EH thanks SERB, Govt. of India, for support through EMR/2015/000622. VR thanks Govt. of India, 
for support through 
DST-INSPIRE/IF170622.


\begin{thebibliography}{99}
\bibitem{snyder}
H. S. Snyder, \textit{Phys. Rev., }\textbf{71}, 38 (1947).
\bibitem{connes}
A. Connes, \textit{Noncommutative Geometry}, (Academic Press, 1994).
\bibitem{seiberg-witten}
N. Seiberg and E. Witten, \textit{JHEP }, \textbf{9909}, 032 (1999).
\bibitem{Nekrasov}
M. R. Douglas and N. A. Nekrasov, \textit{Rev. Mod. Phys., }\textbf{73}, 977 (2001).
\bibitem{Sabo}
R. J. Szabo, \textit{Phys. Rep., }\textbf{378}, 207 (2003).
\bibitem{Glikman}
J. Kowalski-Glikman, \textit{Lect. Notes. Phys., }\textbf{669}, 131 (2005).
\bibitem{Roberts-Freedenhagen}
S. Doplicher, K. Fredenhagen and J. E. Roberts, \textit{Phys. Lett. B., }\textbf{331}, 39 (1994).
\bibitem{am}
J. Ambjorn, J. Jurkiewicz, and R. Loll, \textit{Phys. Rev. Lett., }\textbf{93} 131301 (2004).
\bibitem{madore}
J. Madore, \textit{Class. Quantum Grav., }\textbf{9}, 69 (1992).
\bibitem{bal}
A. P. Balachandran, S. Kurkcuoglu and S. Vaidya, \textit{Lectures on Fuzzy and Fuzzy SUSY Physics, } (World Scientific Publishing Co. Pvt. Ltd., Singapore, 2007).
\bibitem{Minwalla}
S. Minwalla, Van Raamsdonk. M and N. Seiberg N., \textit{JHEP, }\textbf{0002}, 020 (2000). 
\bibitem{Chaichan}
M. Chaichian, P. Kulish, K. Nishijima and A. Tureanu, \textit{Phys. Lett. B., }\textbf{604}, 98 (2004).
\bibitem{Wess}
P. Aschieri, C. Blohmann, M. Dimitrijevic, F. Meye, P. Schupp and J. Wess, \textit{Class. Quantum Grav. }\textbf{22} (2005)
3511; P. Aschieri, M. Dimitrijevic, F. Meyer and J. Wess, \textit{J. Class. Quantum
Grav. }\textbf{23} (2006) 1883.
\bibitem{dsr1}
J. Kowalski-Glikman, \textit{Introduction to Doubly Special Relativity,} (Springer, 2005).
\bibitem{amelio-camelia}
G. Amelino-Camelia, \textit{Phys. Lett. B., }\textbf{510}, 255 (2001).
\bibitem{shanmajid}
S. Majid and H. Ruegg, \textit{Phys. Lett. B, }\textbf{334}, 348 (1994).
\bibitem{lukierski1}
J. Lukierski, A. Nowicki, H. Ruegg, and V. N. Tolstoy, \textit{Phys. Lett. B., }\textbf{264}, 331 (1991).
\bibitem{lukierski2}
J. Lukierski, A. Nowicki, and H. Ruegg, \textit{Phys. Lett. B., }\textbf{293},344 (1992).
\bibitem{lukierski3}
J. Lukierski and H. Ruegg, \textit{Phys. Lett. B., }\textbf{329}, 189 (1994).
\bibitem{mel1}
S. Meljanac and M. Stojic, \textit{Eur. Phys. J. C., }\textbf{47}, 531 (2006).
\bibitem{mel2}
N. Durov, S. Meljanac, A. Samsarov, and Z. Skoda, \textit{Jorunal of Algebra, }\textbf{309}, 318 (2007).
\bibitem{mel3}
S. Meljanac, S. Kresic-Juric, and M. Stojic, \textit{Eur. Phys. J. C., }\textbf{51}, 229 (2007).
\bibitem{mel4}
S. Meljanac, A. Samsarov, M. Stojic, and Kumar S. Gupta, \textit{Eur. Phys. J. C., }\textbf{53}, 295 (2008).
\bibitem{Paschol}
A. Borowiec, K. S. Gupta, S. Meljanac and A. Pachol, \textit{Europhys. Lett., }\textbf{92}, 20006 (2010) 
\bibitem{boro}
A. Borowiec and A. Pachol, \textit{SIGMA, }\textbf{6}, 086 (2010).
\bibitem{pach}
D. Kovacevic, S. Meljanac, A. Pachol and R. Strajn, \textit{Phys. Lett. B., } \textbf{711}, 122 (2012).
\bibitem{jonke}
M. Dimitrijevic, L. Jonke, A. Pachol, \textit{SIGMA, }\textbf{10}, 063 (2014).
\bibitem{moller}
M. Dimitrijevic, L. Jonke and L. Moller, \textit{JHEP, } \textit{086}, 0509 (2005). 
\bibitem{meyer}
M. Dimitrijevic, F. Meyer, L. Moller and J. Wess, \textit{Eur.Phys.J. C., }\textbf{36} (2004).
\bibitem{ytaka}
Y. Takahashi, \textit{An Introduction to Field Quantization, }(Pergamon Press, 1969).
\bibitem{yt} 
Y. Takahashi and H. Umezawa, \textit{Nucl. Phys. } \textbf{51}, 193 (1964).
\bibitem{pal}
Y. Takahashi and R. Palmer, \textit{Phys. Rev. D., } \textbf{10} (1970).
\bibitem{lur}
D. Lurie, Y. Takahashi and H. Umezawa, \textit{J. Math. Phys. }\textbf{7} (1966).
\bibitem{trg}
T. R. Govindarajan, K. S. Gupta, E. Harikumar, S. Meljanac and D. Meljanac, \textit{Phys. Rev. D } \textbf{80}, 025014 (2009).
\bibitem{gup}
T. R. Govindarajan, K. S. Gupta, E. Harikumar, S. Meljanac and D. Meljanac, \textit{Phys. Rev. D } \textbf{77}, 105010 (2008).
\bibitem{eh}
E. Harikumar, \textit{ Eur. Phys. Lett., }\textbf{90}, 21001 (2010).
\bibitem{max}
E. Harikumar, T. Juric and S. Meljanac, \textit{Phys. Rev. D., }\textbf{84}, 085020 (2011).
\bibitem{geo}
E. Harikumar, T. Juric and S. Meljanac, \textit{Phys. Rev. D., }\textbf{86}, 045002 (2012).
\bibitem{kDirac} 
E. Harikumar, M. Sivakumar and N. Srinivas, Mod. Phys. Lett. {\bf A26 }, 1103 (2011).
\bibitem{Ogs}
M. Ostrogradsky, \textit{Mem. Ac. St. Petersbourg, }\textbf{VI 4}, 385 (1850).
\bibitem{braga}
J. Barcelos-Neto and N. R. F. Braga, \textit{Acta Physica Polonica. B., }\textbf{20}, (1989)
\bibitem{bol}
C. G. Bollini and J. J. Giambiagi, \textit{Revista Brasileria de Fisica, }\textbf{17}, 1, (1987).
\bibitem{un}
Unruh, W. G., \textit{Phys. Rev. D., } \textbf{14}, 870–892, (1976).
\bibitem{lcbc}
L. C. B. Crispino, A. Higuchi and  G. A. S. Matsas, \textit{Rev. Mod. Phys., }\textbf{80}, (2008).
\bibitem{bir}
Birrell, N. D.  and P. C. W. Davies, \textit{Quantum Fields in Curved Space, } (Cambridge University Press, 1982).
\bibitem{hck}
H. C. Kim, J. H. Yee and C. Rim, \textit{Phys. Rev. D., }\textbf{75}, 045017 (2007).
\bibitem{ravi1}
E. Harikumar, A. K. Kapoor and Ravikant Verma, \textit{Phys. Rev. D., }\textbf{86}, 045022 (2012).
\bibitem{ravi2}
E. Harikumar and Ravikant Verma, \textit{Mod. Phys. Rev. A., }\textbf{28}, 1350063 (2013).
\bibitem{qdef}
L. C. Bredenharn and M. A. Lohe. \textit{Quantum Group Symmetry and q-Tensor Algebra, } (World Scientific Publishing Co. Pvt. Ltd., Singapore, 1995).
\end{thebibliography}
\end{document}